\documentclass[fleqn,usenatbib]{mnras}

\usepackage{newtxtext,newtxmath}

\usepackage[T1]{fontenc}
\usepackage{ae,aecompl}

\usepackage{graphicx}	
\usepackage{amsmath}	
\usepackage{amssymb}	

\usepackage{comment}

\newcommand{\Ni}{\ensuremath{^{56}\mathrm{Ni}}}
\newcommand{\Co}{\ensuremath{^{56}\mathrm{Co}}}

\newcommand{\Msun}{\ensuremath{\mathrm{M}_\odot}}

\newcommand{\kmps}{\ensuremath{\mathrm{km~s^{-1}}}}

\title[Core collapse for SN~2007bi and SN~1999as]{
Synthetic spectra of energetic core-collapse supernovae and the early spectra of SN~2007bi and SN~1999as
}

\author[T. J. Moriya, P. A. Mazzali, \& M. Tanaka]{
Takashi J. Moriya,$^{1}$\thanks{E-mail: takashi.moriya@nao.ac.jp (TJM)}
Paolo A. Mazzali,$^{2,3}$ and
Masaomi Tanaka$^{4}$ 
\\
$^{1}$Division of Theoretical Astronomy, National Astronomical Observatory of Japan, National Institutes of Natural Sciences, \\
2-21-1 Osawa, Mitaka, Tokyo 181-8588, Japan \\
$^{2}$Astrophysics Research Institute, Liverpool John Moores University,
IC2, Liverpool Science Park, 146 Brownlow Hill, Liverpool L3 5RF, UK \\
$^{3}$Max Planck Institute for Astrophysics, Karl-Schwarzschild-Stra{\ss}e 1,
D-85748 Garching, Germany \\
$^{4}$Astronomical Institute, Tohoku University, 6-3 Aramaki Aza-Aoba, Aoba-ku, Sendai 980-8578, Japan 
}

\date{Accepted 2019 January 22. Received 2019 January 14; in original form 2018 December 6}

\pubyear{2019}

\begin{document}
\label{firstpage}
\pagerange{\pageref{firstpage}--\pageref{lastpage}}
\maketitle

\begin{abstract}
SN~2007bi and SN~1999as are among the first superluminous supernovae discovered. SN~2007bi was suggested to be powered by the radioactive decay of a large amount ($5-10~\Msun$) of \Ni. SN~1999as has a similar spectrum to SN~2007bi. One suggested way to synthesize such a large amount of \Ni\ is through energetic core-collapse supernovae from very massive progenitors. Although the synthetic light curves of extremely energetic core-collapse supernovae have been shown to be consistent with SN~2007bi, no synthetic spectra have been reported. Here, we present synthetic spectra of extremely energetic core-collapse supernovae during the photospheric phases. We find that the ejecta density structure above $13,000 - 16,000~\kmps$ needs to be cut in order to explain the co-existing broad and narrow line absorptions in SN~2007bi and SN~1999as. The density cut is likely caused by the interaction between the supernova ejecta and a dense circumstellar medium. Our results indicate that about 3~\Msun\ of hydrogen-free dense circumstellar media might exist near the progenitors of SN~2007bi and SN~1999as. These massive circumstellar media would significantly affect the light-curve and spectral properties of the supernovae. The precursors that are sometimes  observed in superluminous supernovae might be related to the collision of the ejecta with such dense circumstellar media. We also confirm results of previous studies that synthetic spectra from pair-instability supernova models do not match the early spectra of SN~2007bi and SN~1999as.
\end{abstract}




\section{Introduction}\label{sec:introduction}

Superluminous supernovae (SLSNe) are intrinsically luminous supernovae (SNe)
that have been recognized in the last decade
(\citealt{gal-yam2012slsnreview,gal-yam2018slsnspec,howell2017slsnreview}). They typically reach
magnitudes brighter than $-21$ in optical bands. Broadly speaking,  two kinds of
SLSNe can be distinguished by their spectra -- those with hydrogen lines and
those without. SLSNe with hydrogen lines (Type~II) typically have narrow
emission features that are commonly found in Type~IIn SNe
\citep[e.g.,][]{smith2007sn2006gy,smith2010sn2006gyspectra}, and their huge
luminosity is ascribed to the presence of dense circumstellar media (CSM)
\citep[e.g.,][]{moriya2013sn2006gy,chatzopoulos2013chi2}. SLSNe without hydrogen
lines (Type~I) do not show obvious interaction signatures as  hydrogen-rich
SLSNe do \citep{quimby2011slsnic,quimby2018slsn1spec,howell2013snlsslsn}, and their power sources
are not well understood (see \citealt{moriya2018slsnreview} for a review).

SN~2007bi ($z=0.1279$) was one of the first reported Type~I SLSNe. It reached
$-21.3$~mag in the $R$ band \citep{gal-yam2009sn2007bi,young2010sn2007bi}. It
was estimated to require $5-10~\Msun$ of radioactive \Ni\ to account for the
peak luminosity. Although the rise time is not well constrained, the light-curve
(LC) decline rate is consistent with the decay rate of \Co, which implies that
the LC may be powered by the radioactive decay of \Ni. In addition, synthetic
nebular spectra were found to be consistent with a large production of \Ni\ in a
massive carbon-oxygen core \citep{gal-yam2009sn2007bi}. These facts led
\citet{gal-yam2009sn2007bi} to conclude that SN~2007bi could be a
pair-instability SN (PISN). SN~1999as ($z=0.127$)
\citep{knop1999sn1999as,kasen2004PhDT} is also shown to have similar
spectroscopic properties to SN~2007bi in \citet{gal-yam2009sn2007bi}. The
observations of SN~1999as are summarized in \citet{kasen2004PhDT}.

PISNe are theoretically predicted explosions of very massive stars with helium
core mass between $\sim 70$ and $\sim 140~\Msun$ \citep[e.g.,][]{heger2002pisn}.
The very massive core becomes dynamically unstable owing to the production of
electron and positron pairs during the evolution. The unstable core collapses
and becomes hot enough to trigger explosive oxygen burning. If the core mass is
in the above mass range, explosive oxygen burning can release enough energy to
unbind the whole star, which then explodes. This explosion is called a PISN
\citep[e.g.,][]{rakavy1967pisn,barkat1967pisn}. Explosive nucleosynthsis can
lead to the massive production of radioactive \Ni, enough to explain the large
luminosity of SLSNe. If the core mass is too high, oxygen burning does not
produce enough energy to unbind the whole star, which simply collapses to a
black hole \citep[e.g.,][]{ohkubo2009popiii}.

Although PISNe were originally suggested to account for SN~2007bi and some other
SLSNe, detailed LC and spectral modelling show that PISNe are likely to have
different features from those observed in SLSNe. The LCs of hydrogen-free PISNe
that become as bright as SLSNe are predicted to have rise times of more than
100\,days
\citep[e.g.,][]{kasen2011pisn,dessart2013pisn,chatzopoulos2015rotpisnlcsp,kozyreva2017fastpisn},
while SLSNe tend to have much shorter rise times
(\citealt{nicholl2013ptf12dam,nicholl2015slsndiversity,inserra2013firstmagnetar,decia2018slsnlc,lunnan2018slsnpan},
but see also \citealt{lunnan2016ps1-14bj}). Synthetic spectra of PISNe are found
to be much redder than the observed spectra of SLSNe and they do not match the
observations (e.g.,
\citealt{mazzali2019,dessart2012magni,jerkstrand2016pisnlate}). These issues can
be relaxed if strong mixing in the ejecta occurs in PISNe
(\citealt{mazzali2019,kozyreva2015mixingpisn}), but such a strong mixing is not
found in multidimensional explosion simulations of PISNe
\citep[e.g.,][]{joggerst2011pisnmixing,chatzopoulos2013pisnmultid,chen2014pisnmixing}.
The long rise times and red spectra predicted in luminous PISNe result from the
massive cores required to synthesize a large amount of \Ni, because the core
mass determines the \Ni\ synthesized in PISNe \citep[e.g.,][]{heger2002pisn}.
The requirement of a very massive core to synthesize enough \Ni\ to account for
SLSN luminosity is a fatal issue for the PISN model. It is hard to reconcile the
properties of SN\,2007bi with the current standard picture of PISNe. Therefore,
different power sources than \Ni\ decay, such as spin-down of a strongly
magnetized, rapidly rotating neutron star, a magnetar, have been proposed as a
power source for SN~2007bi \citep[e.g.,][]{kasen2010magnetar,dessart2012magni}.

The massive production of \Ni\ ($5-10~\Msun$) required to explain SN~2007bi,
however, does not necessarily require a PISN explosion. \citet{umeda2008nomoto}
showed that energetic core-collapse SN explosions can produce a large amount of
\Ni\ (up to $\sim 10~\Msun$) if the explosion energy can be as high as
$10^{53}~\mathrm{erg}$. Using their energetic core-collapse SN model,
\citet{moriya2010sn2007bi} showed that the LC of SN~2007bi can be reproduced by
an energetic core-collapse SN explosion of a $43~\Msun$ carbon-oxygen progenitor
exploded with $3.6\times 10^{52}~\mathrm{erg}$ of energy. To synthesize
$5-10~\Msun$ of \Ni, the PISN mechanism requires a carbon-oxygen core of around
100~\Msun\  \citep{heger2002pisn}, while energetic core-collapse SNe can produce
a similar amount of \Ni\ with core masses less than half of that. The
significant reduction in ejecta mass can make the LC rise times significantly
shorter. The similarity of the nebular spectrum of SN~2007bi to that of
SN~1998bw \citep{pm2001}, a broad-line Type~Ic SN associated with the long
gamma-ray burst (GRB) 980425 \citep[e.g.,][]{galama1998sn1998bw}, also indicates
that SN~2007bi may be related to the core-collapse of a massive star
\citep[e.g.,][]{jerkstrand2017longslsnlate,nicholl2018slsnneb}.

The previous study of energetic core-collapse SNe by \citet{moriya2010sn2007bi}
focused on LC modelling, and no spectral modelling was performed. In this paper
we report synthetic spectra of energetic core-collapse SN models during the
early phases, when the photosphere is still in the ejecta, and compare them with
those of SN~2007bi and SN~1999as. The accompanying paper by \citet{mazzali2019}
shows synthetic spectra during the nebular phases from the core-collapse SN model. We also
present some PISN spectral models for comparison.

The rest of this paper is organized as follows. We first introduce our numerical setups and observational data in Section~\ref{sec:setup}. We show our synthetic spectra in Section~\ref{sec:results} and discuss their implications in Section~\ref{sec:discussion}. The conclusions of this paper is summarized in Section~\ref{sec:conclusions}.

\begin{table*}
	\centering
	\caption{Abundance of our SN models. $x$ ($y$) in the table means $x\times 10^{y}$.}
	\label{tab:abundance}
	\begin{tabular}{lcccc} 
		\hline
		element & core-collapse original$^a$ & core-collapse modeling$^b$ & He100$^c$ & He110$^c$ \\
		\hline
	C  & $9.7\ (-3)$ & $9.7\ (-3)$ & $5.6\ (-1)$ & $5.6\ (-1)$ \\
	O  & $7.3\ (-1)$ & $8.3\ (-1)$ & $3.1\ (-1)$ & $2.9\ (-1)$ \\
	Ne & $2.4\ (-2)$ & $2.4\ (-2)$ & $6.1\ (-2)$ & $5.4\ (-2)$ \\
        Mg & $8.6\ (-2)$ & $8.6\ (-2)$ & $5.1\ (-2)$ & $5.2\ (-2)$ \\
        Si & $1.0\ (-1)$ & $1.0\ (-2)$ & $1.6\ (-2)$ & $3.6\ (-2)$ \\
        S  & $3.7\ (-2)$ & $3.7\ (-2)$ & - & $8.1\ (-3)$ \\
        Ar & $5.0\ (-3)$ & $5.0\ (-3)$ & - & $8.9\ (-4)$ \\
        Ca & $4.0\ (-3)$ & $4.0\ (-3)$ & - & $8.9\ (-4)$ \\
        Ti & $2.2\ (-5)$ & $2.2\ (-5)$ & - & - \\
        Fe & $4.8\ (-4)$ & $4.8\ (-4)$ & - & - \\
        Co & $3.0\ (-4)$ & $3.0\ (-4)$ & - & - \\
        Ni & $1.0\ (-5)$ & $1.0\ (-5)$ & - & - \\
        \hline
    \multicolumn{5}{l}{$^a$Average abundance at $7,500-10,000~\kmps$ from the original model in \citet{moriya2010sn2007bi}.}\\
    \multicolumn{5}{l}{$^b$Abundance used in our spectral synthesis for the core-collapse model.} \\
    \multicolumn{5}{l}{$^c$ PISN models.}  
	\end{tabular}
\end{table*}

\section{Model setup}\label{sec:setup}

\subsection{Progenitors}

\subsubsection{Energetic core-collapse SNe}
We adopt the same progenitor and explosion models for SN~2007bi presented in \citet{moriya2010sn2007bi} in this study. The progenitor has a zero-age main-sequence (ZAMS) mass of $100~\Msun$, with metallicity $Z_\odot/200$. Its evolution was numerically followed until core collapse as a single star by \citet{umeda2008nomoto}. The progenitor still has the hydrogen-rich envelope and the helium layer at the time of core collapse. Because SN~2007bi, as well as SN~1999as, is a Type~Ic SN, we artificially remove these outer layers and take the 43~\Msun\ carbon-oxygen core as the SN progenitor. The hydrogen- and helium-rich layers may be lost via binary interaction \citep[e.g.,][]{izzard2004grb,zapartas2017sn2002ap}. Alternatively, a carbon-oxygen core of the same mass may be made by strong mixing in less massive stars \citep[e.g.,][]{yoon2006che,aguilera-dena2018slsngrb}.

\begin{figure}
	\includegraphics[width=\columnwidth]{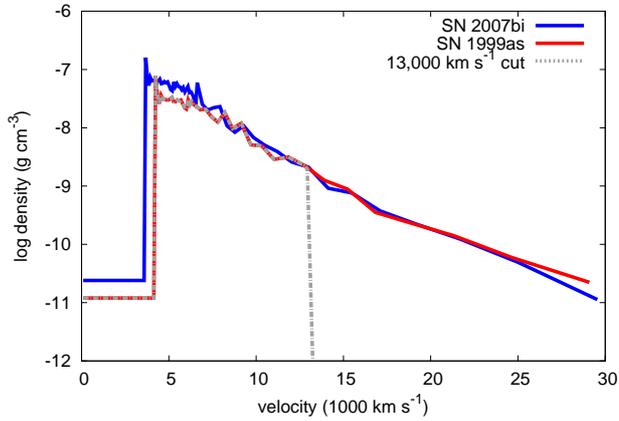}
    \caption{
    Density structure of SN ejecta used for our spectral synthesis of the energetic core-collapse SNe at one day after the explosion. The solid lines are the original density structure without the velocity cut. The dot-dashed density structure is an example of the density structure with the velocity cut at 13,000~\kmps.
    }
    \label{fig:density}
\end{figure}

The explosion is initiated by a thermal bomb, with final explosion energy 
$3.6\times 10^{52}~\mathrm{erg}$ in \citet{moriya2010sn2007bi}. Our spectral
modelling is based on the hydrodynamic structure and the nucleosynthesis
resulting from this explosion model. We set the mass cut at 3~\Msun\ to have
enough amount of \Ni\ to account for the LC of SN~2007bi
\citep{moriya2010sn2007bi}. Thus, the ejecta mass is 40~\Msun.
Figure~\ref{fig:density} shows the density structure at 1~day after the
explosion. The ejecta are homologously expanding at this epoch. The SN~2007bi
model in Figure~\ref{fig:density} is the original model obtained by
\citet{moriya2010sn2007bi} and we use this original density structure for the modelling of SN~2007bi. We find that the spectrum of SN~1999as is better
matched by reducing the ejecta mass to 30~\Msun. Therefore, we use a density
structure scaled to 30~\Msun\ while keeping the explosion energy in modelling
the spectrum of SN~1999as (Figure~\ref{fig:density}). The density structure is
scaled as $v\propto (E_\mathrm{ej}/M_\mathrm{ej})^{0.5}$ and $\rho\propto
(M_\mathrm{ej}^5/E_\mathrm{ej}^3)^{0.5}$, where $E_\mathrm{ej}$ is explosion
energy and $M_\mathrm{ej}$ is ejecta mass. 

\begin{figure}
	\includegraphics[width=\columnwidth]{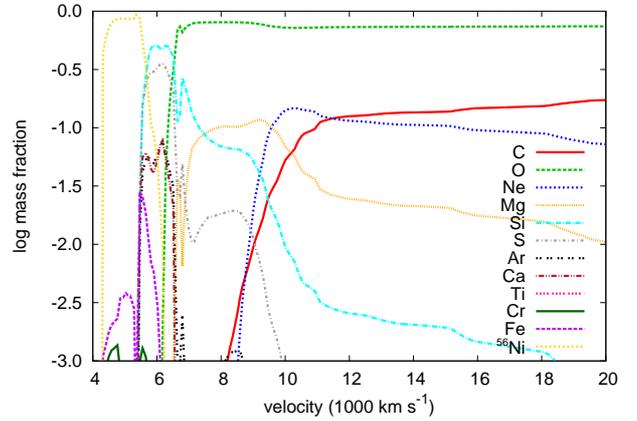}
    \caption{
    Abundance obtained by the explosive nucleosynthesis calculation of \citet{moriya2010sn2007bi}.
    }
    \label{fig:abundance}
\end{figure}

Figure~\ref{fig:abundance} presents the result of the nucleosynthesis calculation by \citet{moriya2010sn2007bi}. The photosphere in our synthetic spectra is located at $7,000-7,500~\kmps$. Thus, the abundance input for our spectral modelling is based on the average abundance between 7,500~\kmps\ and 10,000~\kmps\ (Table~\ref{tab:abundance}).
We found that the synthetic spectra with the original abundance result in too strong silicon lines. Therefore, we artificially reduced the silicon abundance by a factor of 10. The oxygen abundance is increased to compensate.
This density difference may be caused by the mixing of Si-rich and O-rich layers due to the Rayleigh-Taylor instability.
This modified abundance (Table~\ref{tab:abundance}) is used in all the synthetic spectra of energetic core-collapse SNe in this paper. 
Although the ejecta masses of our models for SN~2007bi and SN~1999as are slightly different (40~\Msun\ for SN~2007bi and 30~\Msun\ for SN~1999as), we find that the same abundance above works well for both SNe. We note that the abundance in certain layers where the maximum temperature becomes similar can be similar in energetic massive core collapse SN models \citep{umeda2008nomoto}.

\subsubsection{PISNe}
We have also computed synthetic spectra of PISNe to confirm the results of  previous studies that show that the early spectrum of SN~2007bi is too blue to be a PISN explosion. We take two PISN explosion models computed by \citet{heger2002pisn}, He100 and He110. They are helium stars with metallicity $Z_\odot/100$ at ZAMS whose initial mass is 100~\Msun\ (He100) and 110~\Msun\ (He110). They experience little mass loss during the evolution and they have masses of, respectively, 95~\Msun\ (He100) and 99~\Msun\ (He110) at the time of explosion. Their explosions produce 6~\Msun\ (He100) and 12~\Msun\ (He110) of \Ni, respectively. The LC and spectral properties of these models were previously investigated by \citet{kasen2011pisn,jerkstrand2016pisnlate}. The abundances used in our spectral modelling are listed in Table~\ref{tab:abundance}.

\subsection{Spectral synthesis}
Spectral modelling is performed using the Monte Carlo spectrum synthesis code developed by \citet{mazzali1993lucy,lucy1999mc,mazzali2000mc}. The same code has been used for modelling the spectra of several SLSNe \citep{mazzali2016slsnicsp,chen2016lsq14mo}.

The code assumes the existence of a photosphere and it is suitable for modelling early SN spectra, when the photosphere is in the ejecta. Modeling the nebular spectra requires a different approach and they are studied in the accompanying paper \citep{mazzali2019}. In the spectrum synthesis code we use, blackbody radiation with a given luminosity is emitted from the inner boundary and photon transport in the SN ejecta with a given density structure and composition is solved.
We assume the bolometric luminosity estimated by the observations in our spectral modelling ($5\times 10^{43}~\mathrm{erg~s^{-1}}$ for both SN~2007bi and SN~1999as).
For the PISN models, we use the bolometric luminosity of the models presented in \citet{kasen2011pisn}.

\begin{figure}
	\includegraphics[width=\columnwidth]{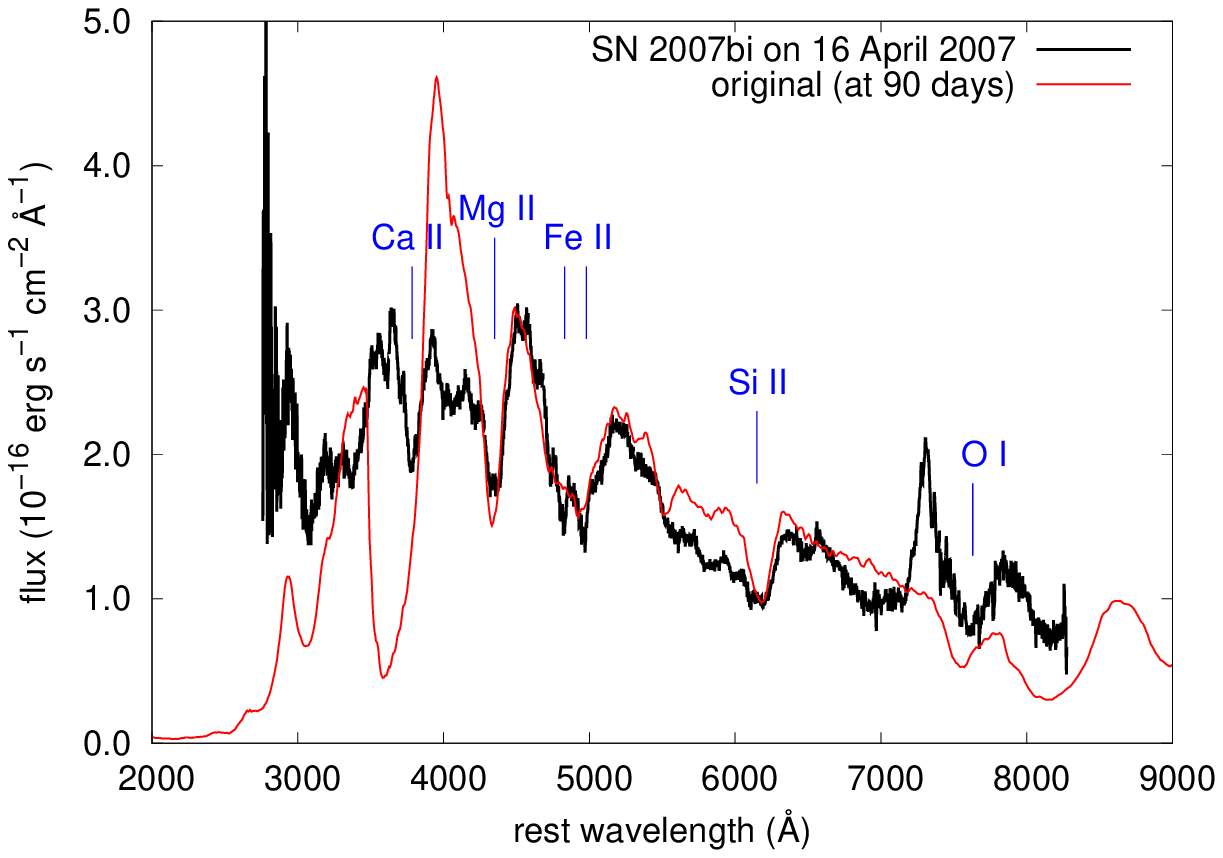}
	\includegraphics[width=\columnwidth]{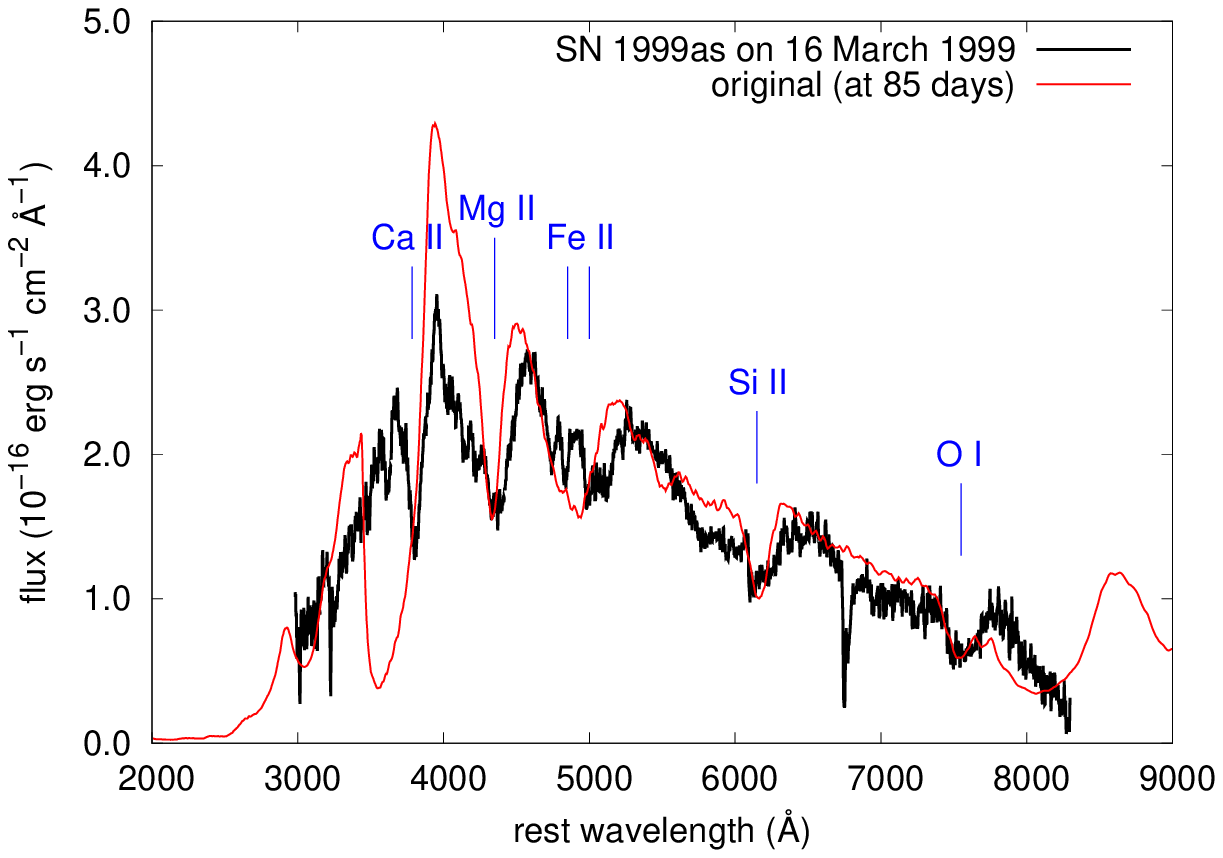}
    \caption{
	Synthetic spectral models with the original density structure without the velocity cut (Figure~\ref{fig:density}).
   }
    \label{fig:spec_original}
\end{figure}

\subsection{Observations}

There are a few spectra during the photospheric phase of SN~2007bi \citep{gal-yam2009sn2007bi,young2010sn2007bi}. We use the earliest observed spectrum, taken 47~days after the LC peak in the rest frame, to compare with our synthetic spectra. As the explosion date of SN\,2007bi is not well constrained, we assume that this spectrum is at 90~days after the explosion. The LC from the energetic core-collapse SN model with mixing in \citet{moriya2010sn2007bi} has a rise time of around 50~days, justifying our assumption on the epoch of the spectrum.

SN\,1999as was the first observed SLSN. The photometric and spectroscopic
observations of SN\,1999as are summarized in \citet{kasen2004PhDT}. We take the
spectrum observed on 16 March 1999, which was shown to have similar properties
to the early spectrum of SN~2007bi \citep{gal-yam2009sn2007bi}. The explosion
date of SN\,1999as is not clear. We assume that 16 March 1999 is 85 days after
the explosion in the rest frame. A slight difference in the assumed epoch does
not affect the overall conclusions in this paper. Some modelling efforts for
SN\,1999as were previously reported in
\citet{hatano2001sn1999as,deng2001sn1999as}.

\section{Synthetic spectra}\label{sec:results}

\subsection{Energetic core-collapse SNe}

We first show the synthetic spectra based on the original density structure without any modifications from the hydrodynamical result (Figure~\ref{fig:spec_original}). The synthetic spectrum of SN~2007bi has the photosphere at 7,000~\kmps\ and the photospheric temperature is 8,300~K. The synthetic spectrum for SN~1999as has the photosphere at 7,500~\kmps\ and the photospheric temperature is 8,200~K. The two synthetic spectra are similar.

The synthetic spectra reproduce key features such as the Ca~\textsc{ii},
Mg~\textsc{ii}, Fe~\textsc{ii}, Si~\textsc{ii}, and O~\textsc{i} lines
(Figure~\ref{fig:spec_original}). However, some lines in the models are broader
than observed. A clear example is the Fe~\textsc{ii} lines at around
$4800-5000$\,\AA. At least two distinct Fe~\textsc{ii} lines are seen in the
observed spectra, while the two lines are merged in the synthetic spectra
because they are too broad to be identified individually.

The breadth of the lines is determined by the velocity extent of the line
forming region. Steeper density slopes make the line forming regions narrower, 
and thus line width decreases with steeper density slopes
\citep[e.g.,][]{mazzali2016slsnicsp,mazzali2017icblsp}. The presence of narrow
absorption lines in the spectra of SN\,2007bi and SN\,1999as suggest that the
density slope in the SN ejecta is steeper than that in our original model.
However, the whole ejecta density slope does not necessarily have to be steeper.
Thus, we make the density structure above a certain velocity steeper than the
original structure to see if the narrow lines can be reproduced. The most
extreme case of a steep density structure is obtained by simply cutting the
density structure above a certain velocity. We focus on this extreme case to
explore above what velocity the ejecta need to have a very steep density
structure, in clear contrast with the original model. Figure~\ref{fig:density}
shows an example of the density structure adopted in the following spectral
modelling. The density structure is assumed to drop suddenly above the specified
velocity cut.

\begin{figure}
	\includegraphics[width=\columnwidth]{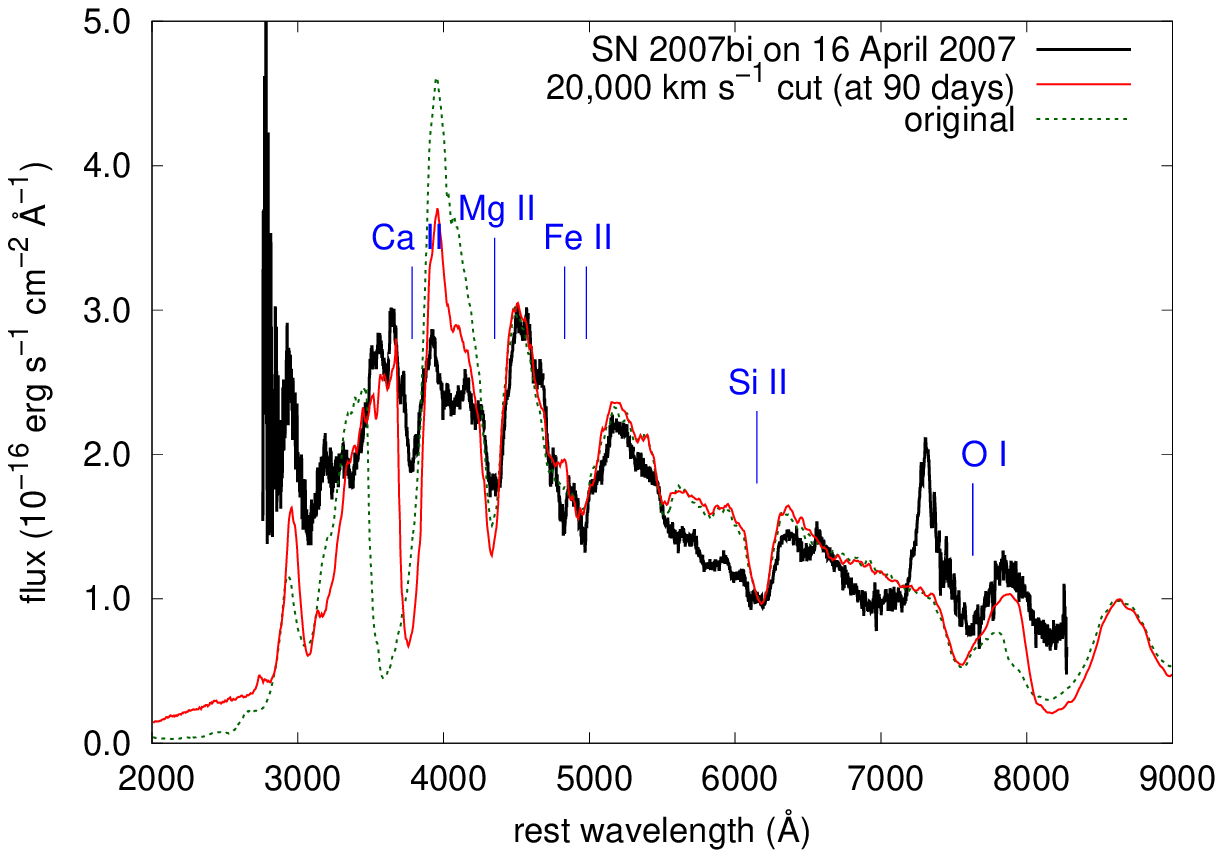}
	\includegraphics[width=\columnwidth]{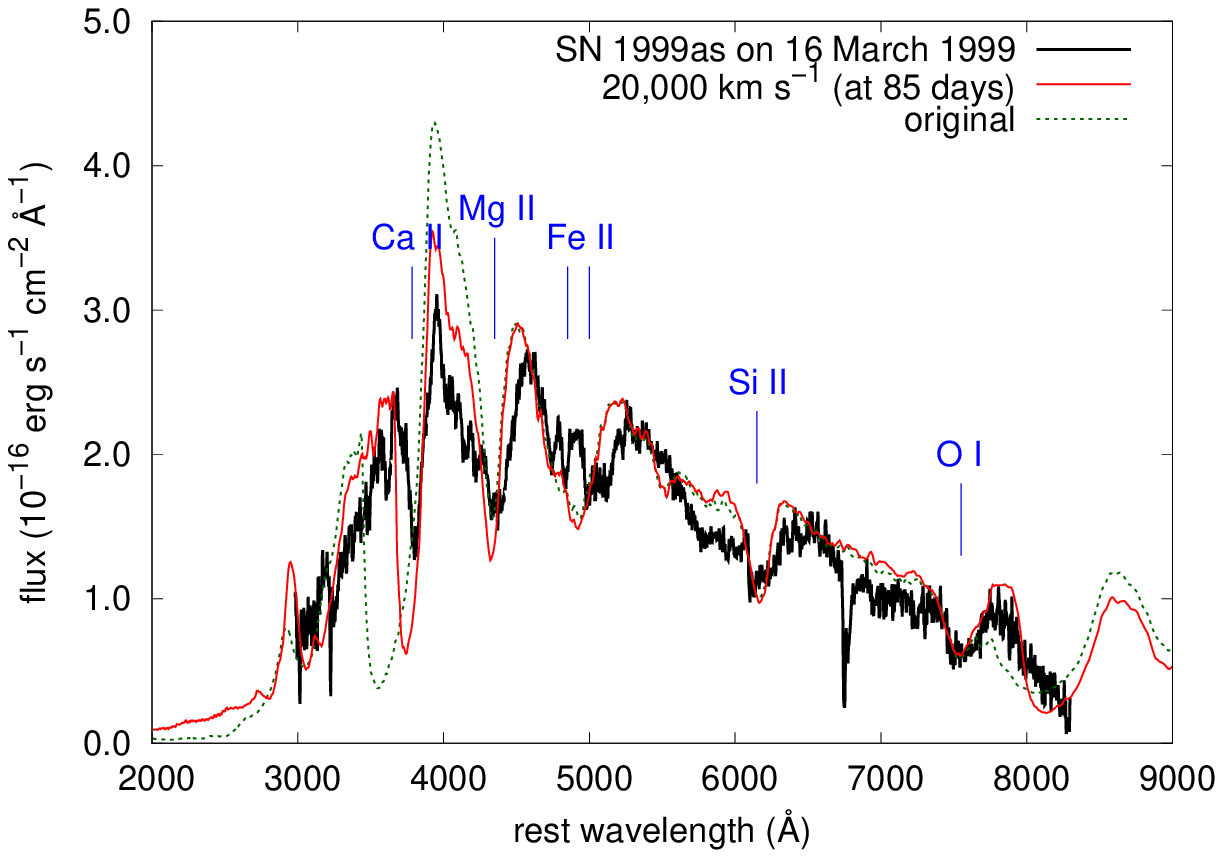}
    \caption{
	Synthetic spectral models from the modified density structure with the velocity cut at 20,000 \kmps.
   }
    \label{fig:spec_20cut}
\end{figure}

Figure~\ref{fig:spec_20cut} shows the synthetic spectra with a velocity cut at
20,000~\kmps. The maximum velocity of the original model reaches near
30,000~\kmps. The mass between 20,000~\kmps\ and 30,000~\kmps\ is 0.9~\Msun\
(SN~2007bi) and 0.6~\Msun\ (SN~1999as). We can clearly see the reduction of the
Ca~\textsc{ii} absorption line width and the weakening of the corresponding
emission strength. We can also see that the narrow Fe~\textsc{ii} lines at
around $4800-5000$~\AA\ are starting to separate, although they are still not as
clearly distinct as observed. The redder side of the spectra is not
significantly affected by the velocity cut, but some improvements are found.

\begin{figure}
	\includegraphics[width=\columnwidth]{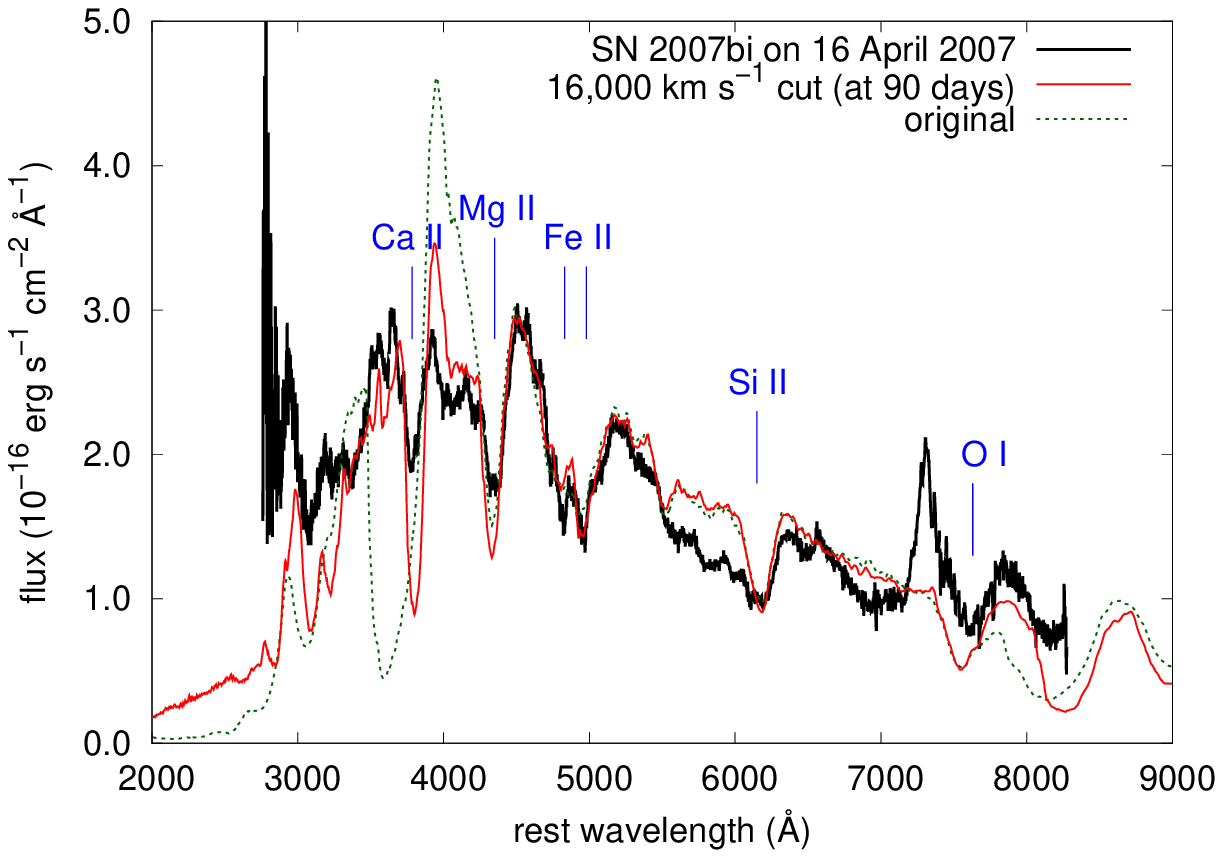}
	\includegraphics[width=\columnwidth]{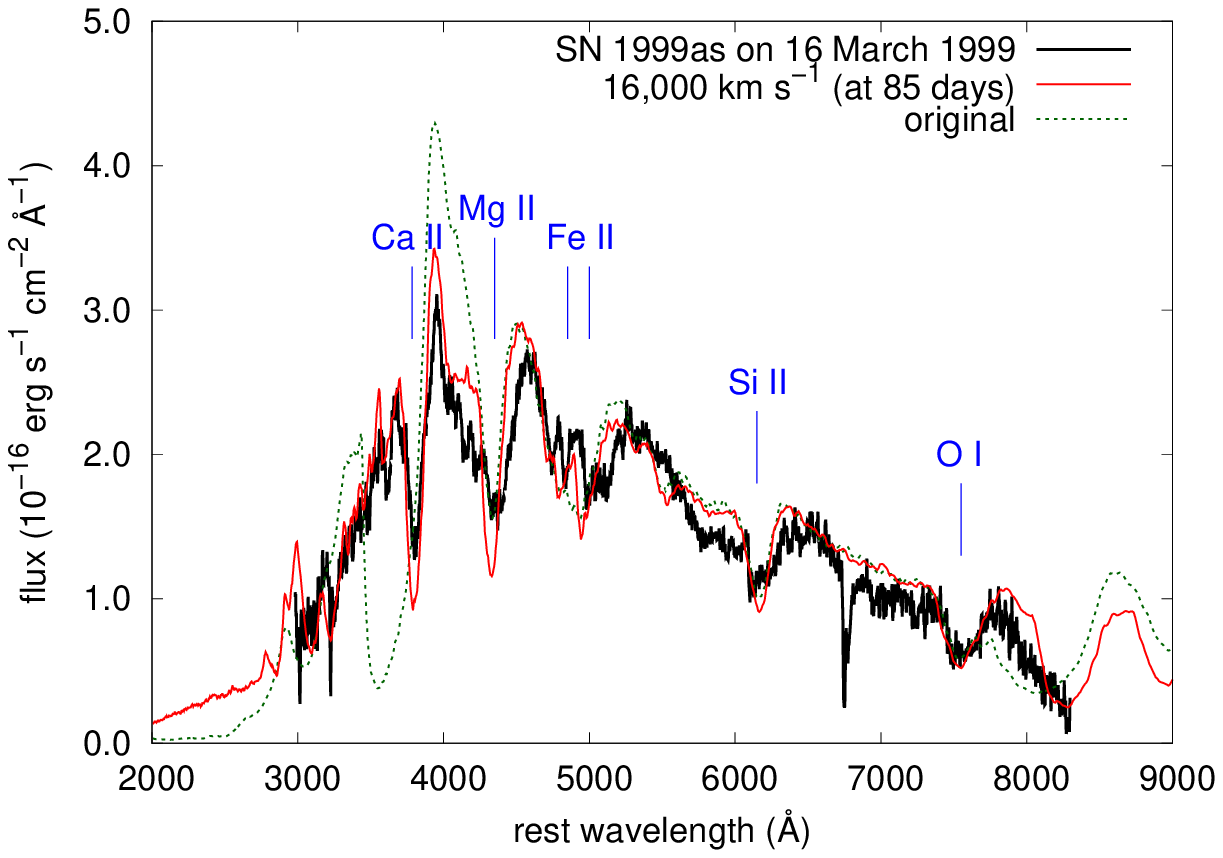}
    \caption{
    Synthetic spectral models from the modified density structure with the velocity cut at 16,000 \kmps. The Fe~\textsc{ii} lines start to clearly separate.
   }
    \label{fig:spec_16cut}
\end{figure}

A clear separation in the Fe~\textsc{ii} lines is found when we set the velocity
cut at 16,000~\kmps\ as presented in Figure~\ref{fig:spec_16cut}. The mass
contained above 16,000~\kmps\ is 2.8~\Msun\ (SN~2007bi) and 3.2~\Msun\
(SN~1999as) and these masses are removed in these models. The line width of the
Ca~\textsc{ii} line is now also well matched. The reduction in the
Ca~\textsc{ii} line width also leads to the appearance of narrow lines of Fe
group elements in the bluer part of the synthetic spectra. Overall, the
synthetic spectra with the 16,000~\kmps\ cut satisfactorily match the observed
spectra.

\begin{figure*}
	\includegraphics[width=\columnwidth]{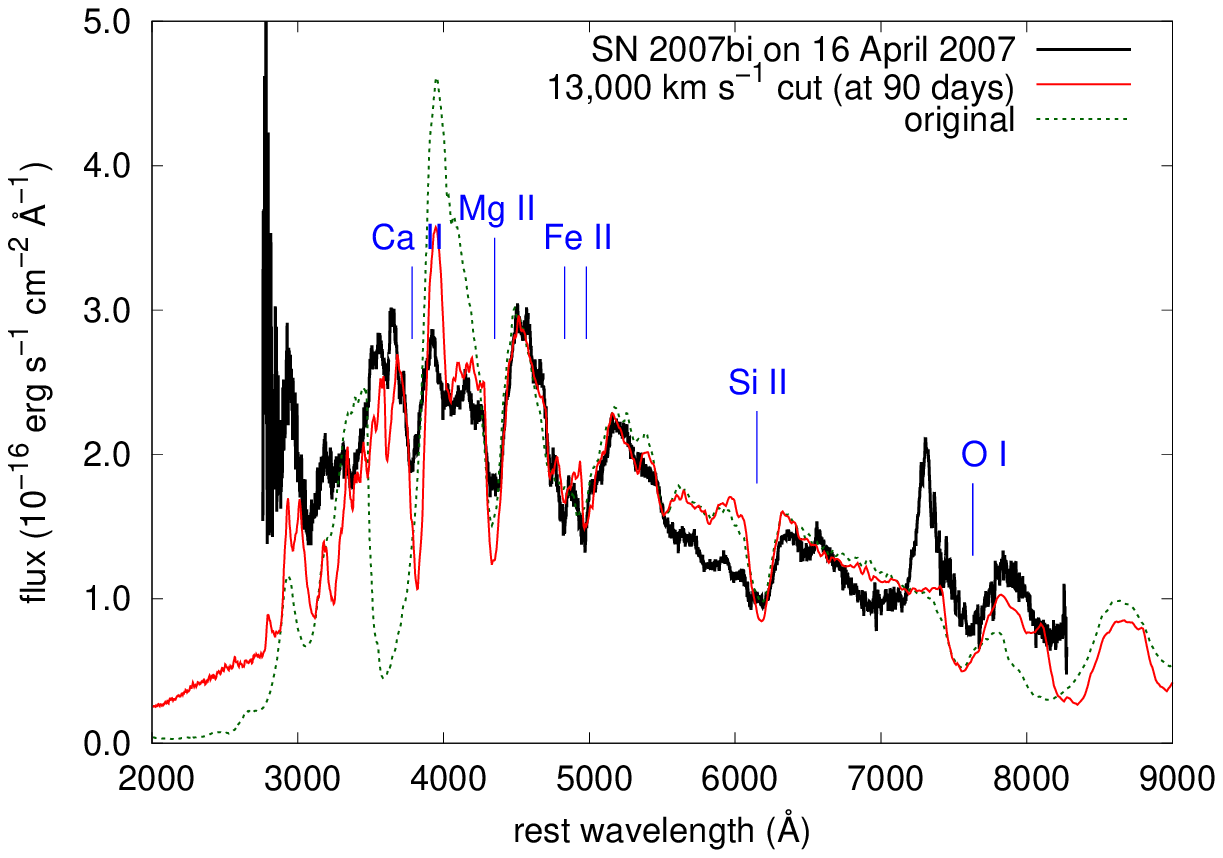}
	\includegraphics[width=\columnwidth]{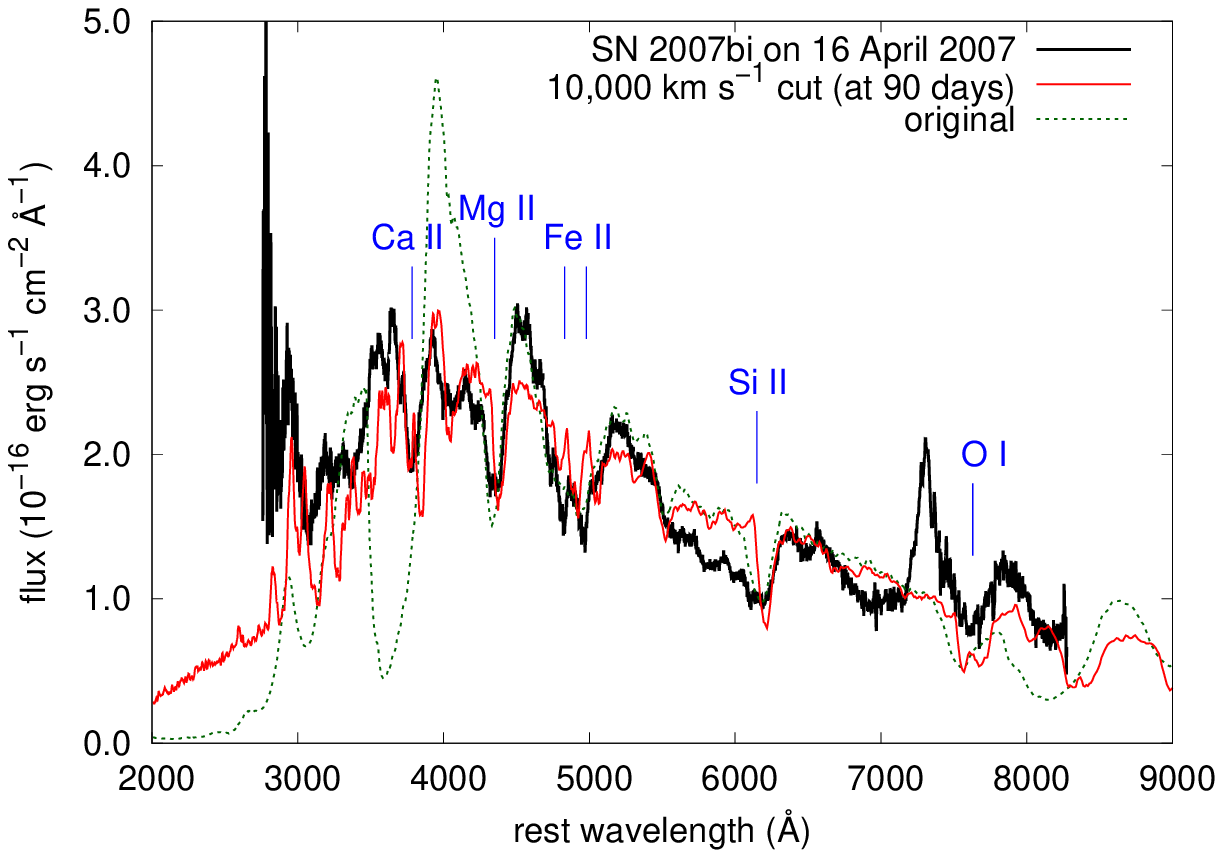} \\
    \includegraphics[width=\columnwidth]{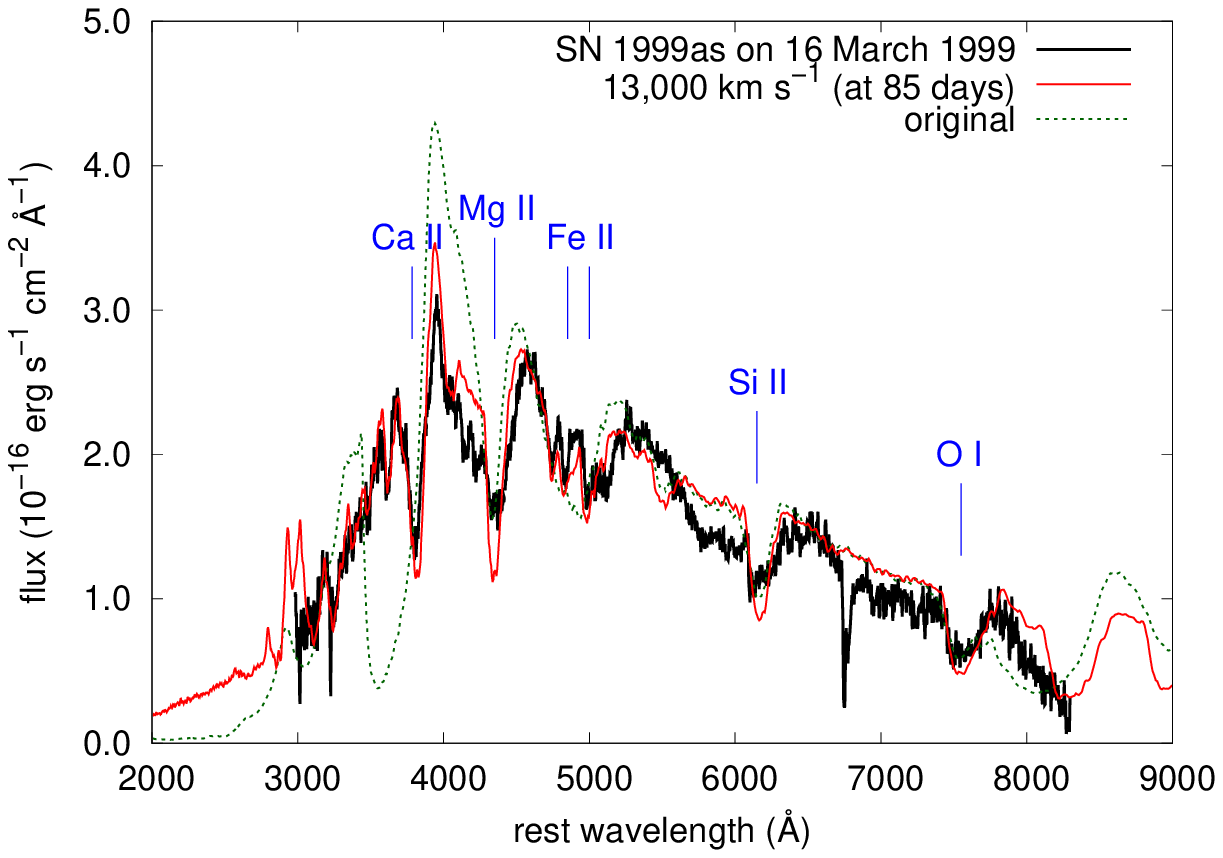}
	\includegraphics[width=\columnwidth]{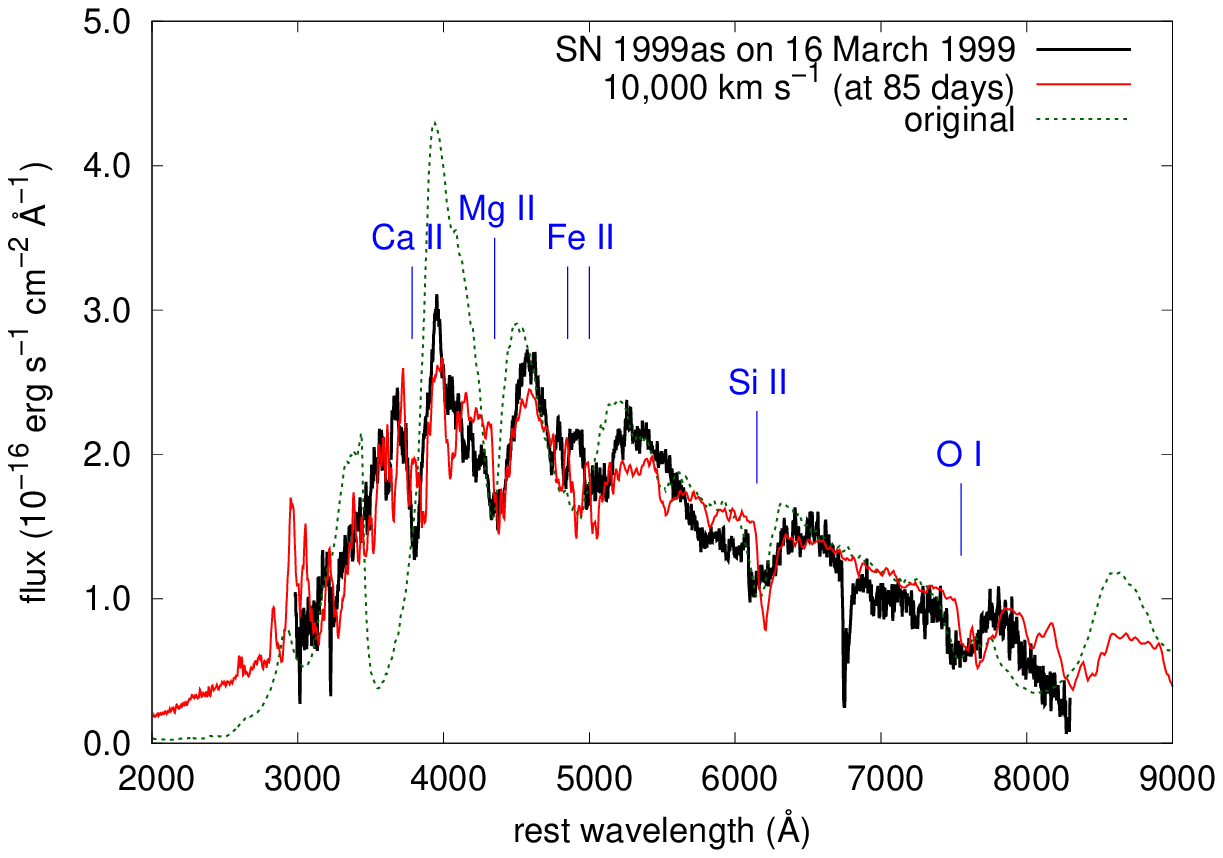}
    \caption{
    Synthetic spectral models from the modified density structure with the velocity cut at 13,000~\kmps\ (left) and 10,000~\kmps\ (right).
   }
    \label{fig:spec_1310cut}
\end{figure*}

Figure~\ref{fig:spec_1310cut} shows the synthetic spectra with velocity cuts at
13,000\,\kmps\ (5.1~\Msun\ cut in SN\,2007bi and 5.6~\Msun\ cut in SN\,1999as)
and 10,000\,\kmps\ (10.5~\Msun\ cut in SN\,2007bi and 10.6~\Msun\ cut in
SN\,1999as). The models cut at 13,000\,\kmps\ reproduce the narrow features in
SN\,2007bi and SN\,1999as, and they are overall as good as the models cut at 
16,000\,\kmps\ presented in Figure~\ref{fig:spec_16cut}. In the models cut at
10,000\,\kmps\ the synthetic spectra start to show too many narrow features that
are not observed in SN\,2007bi and SN\,1999as. Thus, the velocity cut should be
at a velocity larger than 10,000\,\kmps. 

Given the above results, we conclude that the synthetic spectra from the
energetic core-collapse SN model are consistent with the observed spectral
features of SN~2007bi and SN~1999as, including both broad and narrow components,
if we introduce a velocity cut between 16,000~\kmps\ and 13,000~\kmps. We
discuss the origin of the velocity cut in Section~\ref{sec:discussion}.

\subsection{PISN models}

\begin{figure}
	\includegraphics[width=\columnwidth]{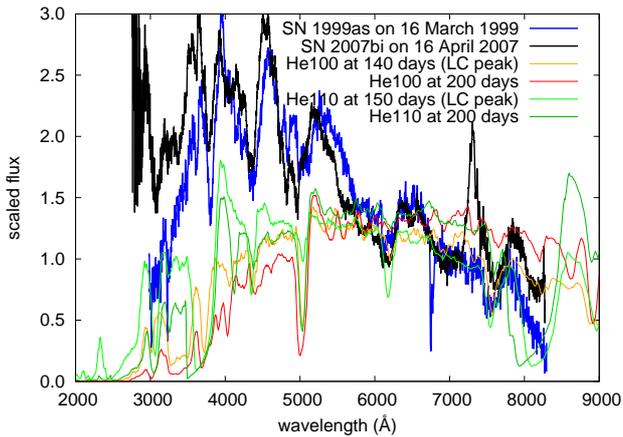}
    \caption{
    Comparison between the synthetic PISN spectra and the spectra of SN~2007bi and SN~1999as.
    }
    \label{fig:pisn}
\end{figure}

Figure~\ref{fig:pisn} presents our synthetic PISN spectra and their comparison
with the spectra of SN~2007bi and SN~1999as. Because the spectra of SN~2007bi
and SN~1999as were observed after the LC peak, we show the synthetic spectra at
and after the LC peak. The PISN spectra at LC peak are already much redder than
SN~2007bi and SN~1999as, confirming the results of previous studies that
SN~2007bi, as well as SN~1999as, have too blue spectra to be PISNe
\citep[e.g.,][]{dessart2012magni}.

\section{Discussion}\label{sec:discussion}

We have shown that the energetic core-collapse SN model of
\citet{moriya2010sn2007bi} can reproduce the overall spectral features of
SN~2007bi and SN~1999as if a velocity cut at around $16,000-13,000~\kmps$ in the
density structure is introduced. With the velocity cut, both broad and narrow
spectral lines are reproduced. Such narrow absorptions are sometimes found in
SLSN spectra \citep{liu2017slsnmodj,quimby2018slsn1spec,gal-yam2018slsnspec}. They can be
explained by the existence of a velocity cut in the ejecta.

Such a density cut in the ejecta might be related to the density structure of
the progenitor. Alternatively, we speculate that a promising physical origin of
the velocity cut is the deceleration of the outermost layers in the SN ejecta by
the collision with an external dense CSM. If a dense CSM surrounds the
progenitor, the outer layers can be decelerated and a slow, dense and cool shell
is formed between the SN ejecta and the dense CSM
\citep[e.g.,][]{moriya2013sn2006gy,sorokina2016slsncsm}. In the models of both 
SN\,2007bi and SN\,1999as, the mass in the ejecta above 16,000\,\kmps\ is about
3\,\Msun\ and a similar amount of dense CSM is required to exist to make the
velocity cut and account for the narrow spectral lines.

\citet{kasen2004PhDT} previously proposed a similar idea to explain the narrow
spectral features observed in SN\,1999as. Unlike this paper,
\citet{kasen2004PhDT} did not use a hydrodynamical model for the ejecta
structure and did not perform LC modelling as we did in our previous study
\citep{moriya2010sn2007bi}. However, his simplified modelling led to the
conclusion that a velocity cut at $12,500-14,000\,\kmps$ can explain the narrow
spectral features observed in SN\,1999as, which is in full accord with our
conclusion.

The kinetic energy contained in the outermost layers above 16,000\,\kmps\ is
$\sim 10^{52}\,\mathrm{erg}$ in both SN~2007bi and SN~1999as models. If the
outermost layers are decelerated, this huge amount of energy could be released
as radiation and the SN luminosity may have been enhanced by the extra light
from interaction at early times. Thus, all the luminosity in SN\,2007bi may not
necessarily come from \Ni\ decay as proposed by \citet{moriya2010sn2007bi}.
Still, the consistency of the late-time spectrum and photometry of SN\,2007bi
with \Ni\ decay energization cannot be discounted.
\citet{tolstov2017ppisn,tolstov2017gaia16apdcsm} also propose that more luminous
SLSNe such as iPTF12dam and Gaia16apd are powered by the interaction between
energetic core-collapse SNe and massive dense CSM, although they found more than
10\,\Msun\ of hydrogen-free CSM is required to explain these SLSNe. In this case,
however, we do not expect to see the absorption line spectrum that is typical of
Type~Ic SLSNe, making hard for this scenario to reconcile with the observations.
The precursor bump observed in SLSNe
\citep[e.g.,][]{leloudas2012sn2006oz,nicholl2016slsnbump} could also be related
to such a deceleration of the outer layers \citep{moriya2012dip}, making the
subsequent narrow spectral features. A Type\,Ic SN\,2010mb had an early luminosity excess which is attributed to the interaction between the SN ejecta and dense CSM \citep{ben-ami2014sn2010mb}. Interestingly enough, the estimated CSM mass for SN~2010mb ($\sim 3$\,\Msun) is similar to those estimated for SN\,2007bi and SN\,1999as here.

The early interaction between the SN ejecta and dense CSM would result in the formation of the cool dense shell surrounding the SN ejecta. 
The cool dense shell contains $\simeq 3-5~\Msun$ which is about 10\% of the original ejecta and the diffusion time in the ejecta does not change much by losing this small amount. Thus, the main part of the LC as presented in \citet{moriya2010sn2007bi} is not affected much by the existence of the dense CSM.
Our original hydrodynamic model does not take the effect of the dense CSM into account and the cool dense shell does not exist in the density structure we use for the spectroscopic modelling. The formation of the cool dense shell prevents the reverse shock to propagate quickly into the ejecta and the hydrodynamic structure below the dense cool shell would not be affected much by the possible CSM interaction we propose \citep[e.g.,][]{moriya2013sn2006gy}. Therefore, our simple density cut would be a reasonable assumption.
Still, the dense shell does exist above the photosphere and it may have a certain effect the spectroscopic properties. Because the cool dense shell is unstable, it is likely that the shell is deformed and creates a clumpy structure. The effect of the clumpy structure needs to be investigated \citep[e.g.,][]{chugai1994danziger}, but this is beyond the scope of this paper.

The existence of hydrogen-free dense CSM in Type~I SLSNe has been speculated in
many studies
\citep{chevalier2011irwin,moriya2012dip,ginzburg2012slsncsm,chatzopoulos2013chi2,sorokina2016slsncsm,chen2016lsq14mo,tolstov2017ppisn,tolstov2017gaia16apdcsm}.
The existence of a CSM shell is also identified in the Type~Ic SLSN iPTF16eh
\citep{lunnan2018shell}, as well as in three Type~Ic SLSNe with late-phase
hydrogen emission (\citealt{yan2015iptf13ehe,yan2017slsniclateh}, but see also
\citealt{moriya2015slsnhstri}). In our energetic core-collapse SN models, the
dense CSM may have been formed shortly before the explosion by the pulsational
pair-instability \citep{woosley2007sn2006gy,woosley2017ppisn} because the
carbon-oxygen core of our progenitor is massive enough to initiate the
instability. Another possible mechanism to form such a massive CSM shortly
before the explosion is through the rapid progenitor contraction that is linked
to the final progenitor evolution \citep{aguilera-dena2018slsngrb}. The exact
mechanism to form a massive CSM remains a mystery, and further studies are
needed to identify it.

The carbon-oxygen core mass of around 40~\Msun\ required to reproduce SN~2007bi
is close to the maximum core mass predicted to exist. \citet{woosley2017ppisn}
shows that carbon-oxygen cores above 40~\Msun\ suffer from the pulsational pair
instability and a part of the core is inevitably ejected. For example, a
48~\Msun\ helium star with a carbon-oxygen core of 40~\Msun\ loses 7~\Msun\ from
the surface shortly before core collapse, leaving only 41~\Msun\ at the time of
core collapse \citep{woosley2017ppisn}. This combination of core and CSM mass
matches well what is needed for our energetic core-collapse SN picture for
SN~2007bi. \citet{woosley2017ppisn} shows that the carbon-oxygen core mass is
limited to below around 45~\Msun\ at core collapse. Therefore, it may be 
difficult to explain Type~Ic SLSNe requiring ejecta exceeding around 50~\Msun\
as energetic core-collapse explosions.

One remaining question is how to achieve the huge explosion energy ($\simeq 4\times 10^{52}~\mathrm{erg}$) required to synthesize 6~\Msun\ of \Ni\ for SN~2007bi. This amount of energy needs to be provided in a short timescale in the massive core in order to make the temperature high enough to have  sufficient \Ni\ synthesis \citep[e.g.,][]{woosley2002rev,umeda2008nomoto}. The neutrino-driven explosion mechanism cannot provide this amount of energy \citep[e.g.,][]{janka2012ccrev}. A candidate energy source is the quick spin down of a strongly magnetized rapidly rotating neutron star (magnetar). Depending on the equation of state of nuclear matter, the magnetar can have an initial rotational energy of as high as $10^{53}~\mathrm{erg}$ \citep[cf.][]{metzger2015magdiv}. A GRB and associated SN in total have an explosion energy of $(1-2)\times 10^{52}~\mathrm{erg}$, which indicates that their major power source could be the magnetar spin down \citep{mazzali2014grb}. It is possible that SN~2007bi originates from the spin down of a magnetar that happened to acquire more rotational energy through its evolution than those of GRB progenitors.

\section{Conclusions}\label{sec:conclusions}

We presented synthetic spectra of massive ($40~\Msun$), energetic ($3.6\times
10^{52}~\mathrm{erg}$) core-collapse SNe during the photospheric phase. 
Energetic core-collapse SN explosions are suggested to account for SLSNe with
slow LC declines such as SN\,2007bi \citep{moriya2010sn2007bi}. A previous study
by \citet{moriya2010sn2007bi} showed that the LC of SN\,2007bi is consistent
with the energetic core-collapse SN explosion model. However, no spectroscopic
modelling had been performed for this model. We have performed a spectrum
synthesis calculations during the photospheric phase and compared the results
with SN\,2007bi as well as SN\,1999as, a SLSN with properties similar to
SN\,2007bi. In the accompanying paper, \citet{mazzali2019} investigate the
nebular spectral properties of the massive energetic core-collapse SN model.

We have found that the synthetic spectra from the energetic core-collapse SN
model match the early spectra of SN~2007bi and SN~1999as if the density
structure above $13,000-16,000\,\kmps$ is cut. The velocity cut is required in
order to have narrow spectral line components in addition to the broad ones
observed in these SNe. We have also calculated spectra of PISN models and
confirm previous results that these models produce spectra that are too red to
match those of SN\,2007bi.

A possible origin of the velocity cut is the existence of massive CSM around the progenitor. The mass in the ejecta above the velocity cut ($13,000-16,000~\kmps$) is $\simeq 3-5~\Msun$. If a similar amount of CSM exists around the progenitor, the outer ejecta layers can be decelerated, causing what is seen as a velocity cut in the density structure. Because the outer layers contain $\sim 10^{52}~\mathrm{erg}$, their deceleration can provide a part of the SLSN luminosity, especially at early phases. We speculate that the precursor often observed in SLSNe may be related to the deceleration of such an outer layer. We speculate that the interaction between the dense CSM and SN ejecta may result in the formation of a dense clumpy CSM around the SN ejecta, which may affect the spectroscopic properties.



\section*{Acknowledgements}
This work was achieved using the grant of NAOJ Visiting Joint Research supported by the Research Coordination Committee, National Astronomical Observatory of Japan (NAOJ), National Institutes of Natural Sciences (NINS).
TJM is supported by the Grants-in-Aid for Scientific Research of the Japan Society for the Promotion of Science (16H07413, 17H02864, 18K13585).




\bibliographystyle{mnras}
\bibliography{reference} 







\bsp	
\label{lastpage}
\end{document}